\documentclass[conference]{IEEEtran}
\usepackage{cite}
\usepackage{amssymb}
\usepackage{cite}
\usepackage{xcolor}
\usepackage[noend]{algpseudocode}
\usepackage{multirow,booktabs}
\usepackage{algorithm}

\ifCLASSINFOpdf
   \usepackage[pdftex]{graphicx}
\else
\fi
\usepackage{array}
\usepackage{amsmath}
\begin{document}
%
\title{Tree-Chain: A Fast Lightweight Consensus  Algorithm for  IoT Applications}

\author{\IEEEauthorblockN{Ali Dorri}
\IEEEauthorblockA{School of Computer Science\\
Queensland University of Technology (QUT)\\
 ali.dorri@qut.edu.au}
\and
\IEEEauthorblockN{Raja Jurdak}
\IEEEauthorblockA{School of Computer Science\\
	Queensland University of Technology (QUT)\\
	 r.jurdak@qut.edu.au}
}

\maketitle

%
\IEEEpeerreviewmaketitle

\begin{abstract}
	
	Blockchain has received tremendous attention in non-monetary applications including the Internet of Things (IoT) due to its salient features including decentralization, security, auditability, and anonymity. Most conventional blockchains rely on computationally expensive consensus algorithms, have limited throughput, and high transaction delays. In this paper, we propose tree-chain a scalable fast blockchain instantiation that introduces two levels of randomization among the validators: i) transaction level where the validator of each transaction is selected randomly based on the most significant characters of the hash function output (known as consensus code), and ii) blockchain level where validator is randomly allocated to a particular consensus code based on the hash of their public key. Tree-chain introduces parallel chain branches where each validator commits the corresponding transactions in a unique ledger. Implementation results show that tree-chain is runnable on low resource  devices and incurs low processing overhead, achieving near real-time transaction settlement.
	
\end{abstract}


\section{Introduction}\label{sec:intro}
Blockchain, the enabling technology of Bitcoin, has received tremendous attention in recent years due to its salient features including security, anonymity, auditability, trust, transparency, and decentralization.  Blockchain is part of a broader technology, known as Distributed Ledger Technology (DLT) where information is grouped in the form of blocks and the participating nodes reach agreement over the state of the database by following a consensus algorithm.  The latter ensures that every block in the chain is valid,  prevents any single entity from controlling the entire blockchain, and introduces randomness and unpredictability among the nodes that append blocks in the blockchain, also known as validators. \par 

The salient features of blockchain made it attractive for large-scale distributed networks such as the Internet of Things \cite{atlam2018blockchain,christidis2016blockchains,alphand2018iotchain}. However, applying blockchain in IoT is not straightforward as most consensus protocols are computationally demanding which are not necessarily suited for IoT with millions of heterogenous resource-restricted devices \cite{dorri2019lsb,ramachandran2018blockchain,qiu2018dynamic}. With the widespread use of blockchain in a range of diverse domains, multiple consensus algorithms have been proposed to reduce the overheads and fit the specific needs of the target application. However, the existing consensus algorithms suffer from  limited throughput, resource consumption, lack of efficiency, delay in storing transactions, and  overhead in retrieving transactions. Additionally,  the existing  consensus algorithms  may lead to centralization as the node with the highest mining power may be able to control the network, e.g., in Bitcoin mining pools may eventually collude to control the ledger. \par 

\begin{figure}
	\begin{center}
		\includegraphics[width=8cm ,height=8cm ,keepaspectratio]{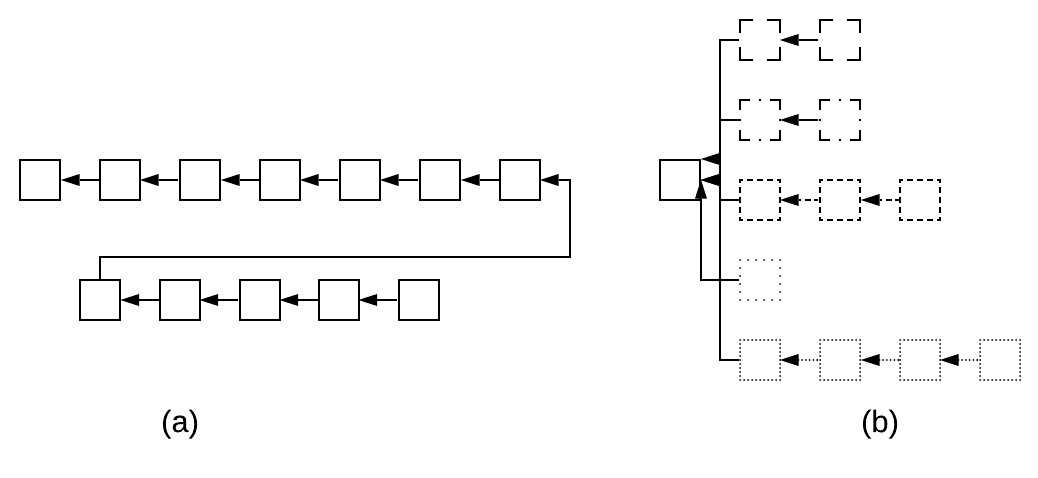}   
		\caption{A highlevel view of a) conventional blockchains, b) tree-chain (blocks with the same border share the same consensus code).}
		\label{fig:chains}
	\end{center}
\end{figure}

To address the aforementioned challenges, in this paper, we propose Tree-chain that bases validator selection on an existing random function in virtually all blockchains: the  hash function output. As shown in Figure \ref{fig:chains}.a, in conventional blockchains, all validators chain their transactions to a  single valid ledger known as the longest ledger. However, this reduces the blockchain efficiency as it wastes computational resources of the validators whose block is not stored in the blockchain (see Section \ref{sec:literature-review}), limits blockchain throughput, and increases delay in storing transactions in the blockchain. Tree-chain, as shown in Figure \ref{fig:chains}.b, consists of multiple parallel chains where in each chain a single validator commits transactions whose content   hash  starts with particular characters, referred to  as consensus code, without the need to dedicate computational resources for consensus. While all validators  still store the entire chain, tree-chain randomizes  how validators commit content at two levels: transaction and blockchain levels.  \par 

At the transaction level, the validator to store each transaction is identified arbitrarily based on the hash value of the transaction content which is random. At the blockchain level, each validator is allocated to a particular consensus code based on the hash of the validator Public Key (PK). Each  character in the hash of a PK corresponds to a particular numeric weight defined in a  dictionary. For each PK, the validators calculate a Key Weight Metric  (KWM) by adding the weights of the symbols in the hash of the PK. Each validator calculates KWM for all PKs of potential validators as well as its own PK. The results are then ordered in descending order. The consensus codes are allocated to the validators in order starting from the largest KWM. \par

Each validator continuously stores blocks in the blockchain for a particular duration of time known as epoch time. At the end of each epoch time, the validators choose a new PK and repeat the outlined algorithm to ensure the consensus code corresponding to each validator changes in each epoch time. Tree-chain does not require the validators to solve any puzzle before appending new blocks, thus the transactions can be stored in the blockchain with negligible delay which makes it appropriate for real-time applications of IoT. The upper bound throughput is the speed at which the validator can verify transactions, group them in blocks, and append the block to the blockchain.   Essentially the computational cost of tree-chain is the KWM computation and ordering of all KWMs, which is significantly lower than the computational cost of the current consensus algorithms.  As outlined earlier in this section, throughput is one of the fundamental challenges in applying blockchain for IoT. Tree-chain is self-scaling as it can adjust blockchain throughput in response to an increase in transaction load. Recall that each validator stores transactions with a particular consensus code. If the number of transactions with a particular code increases, the corresponding validator randomly selects a new validator from an interested pool of non-validator nodes. To ensure randomness among the interested nodes, the node with the highest KWM is chosen as the new validator.  The original consensus code is then split into two smaller range consensus codes, where one is assigned to the original validator and the other to the new validator. Tree-chain is lightweight from a computation perspective, while achieving high security which makes it suitable for both public and private blockchains in large scale networks such as  IoT.  \par

The rest of the paper is organized as follows. Section \ref{sec:literature-review} studies the existing  solutions to reduce blockchain overheads.  Section \ref{Sec:HashCons} outlines the details of tree-chain. Section \ref{sec:security-analysis} provide an analysis of the security of tree-chain and Section \ref{sec:performance} studies the performance of tree-chain. Section \ref{sec:discussions} discusses the future research directions and finally Section \ref{sec:conclusion} concludes the paper.

\section{Literature Review}\label{sec:literature-review}
In this section, we review  the existing works relevant to Tree-chain.  We first discuss the   existing blockchain-based solutions for IoT in Section \ref{sub:sec:lightweight-chain} followed by a review on the existing consensus algorithms in Section \ref{sub-sec-consensus}.  

\subsection{Blockchain-based IoT }\label{sub:sec:lightweight-chain}

In recent years, blockchain applications in IoT has received tremendous attention from academia and practitioners. The authors in \cite{ma2019privacy} proposed a hierarchical blockchain-based   access control in IoT that consists of three  layers: i) device layer: this layer comprises of  IoT devices, ii) fog layer: this layer comprises of  higher resource available devices that connect the IoT to the blockchain, and iii) cloud layer: this layer comprises of cloud servers that manage the blockchain by verifying and  appending new blocks.  In \cite{dorri2017blockchain} the authors studied the blockchain applications to secure  communications among the smart vehicles. The authors in \cite{mengelkamp2018blockchain}  studied the blockchain applications in managing smart grids. An energy marketplace framework is proposed where the energy consumers and producers can trade energy without relying on trusted third parties. The authors in \cite{lee2017blockchain} proposed a blockchain-based solution to remotely update the software of IoT devices. The framework ensures security of communications and the exchanged software update which in turn protects against modified software updates. The authors in \cite{jo2018hybrid} proposed a blockchain-based solution to share health data in a secure, reliable and private manner. The proposed framework is a hierarchical approach where only authorized nodes can access data of the patients.

Due to the significant potential of blockchain, multiple blockchain instantiations have been proposed by academia and practitioners to adopt blockchain based on the requirements of non-monetary applications.  Ethereum \cite{wood2014ethereum}  is a blockchain framework introduced in 2014 that enables the blockchain participants to run Distributed Applications (DApps)  on top of the blockchain. Based on the computational resources demanded by each DApp, the user must pay a fee to the blockchain participants who run the code.  Hyperledger \cite{Hyperledger}  is a project run by Linux foundation that comprises of a number of blockchains each optimized for particular applications. Hyperledger Fabric  \cite{HyperledgerFabric} is run by IBM and aims to provide blockchain solutions for industry applications. The consensus algorithm (see Section \ref{sub-sec-consensus})  can be plugged in based on application requirements which provides high flexibility. 

The authors in \cite{tomescu2017catena} proposed a framework where a summary of a group of transactions is stored in the blockchain to reduce the blockchain memory footprint and increase throughput. A logging server collects transactions and forms them into a single log transaction that essentially contains the hash of each transaction. The latter is then stored in the blockchain. In \cite{liu2019mathsf} the authors proposed a lightweight blockchain instantiation known as LightChain. LightChain encourages the IoT nodes to collaborate by defining a collaboration index that impacts the mining power of a node. To reduce the size of the blockchain an unrelated block offloading filter is introduced that offloads the old blocks and thus not all nodes require to store those.  In our previous work \cite{dorri2019lsb} we proposed a lightweight scalable blockchain (LSB)  for IoT ecosystem.  LSB introduces a time-based consensus algorithm that allows the validators to generate one block per pre-defined time intervals. LSB introduces a throughput management algorithm to ensure self-scaling feature of blockchain.

In conventional blockchains all transactions are boradcast and verified by all the participating nodes which in turn increases the packet and computational overheads. To enhance the blockchain scalability, the concept of \textit{sharding} is proposed in the literature \cite{luu2016secure} that refers to partitioning the network into different groups, i.e., shards, where the nodes in each shard are only responsible to manage transactions in their own shard. The information of each shard is shared with all other shards enabling decentralized management of the blockchain, however, only the nodes in each shard verify and store transactions in the corresponding shard.  

In this section, we studied the blockchain applications in IoT. Consensus is the key to the blockchain that impacts the computational overhead, delay and throughput. The main contribtuon of tree-chain is to introduce a fast and  lightweight consensus algorithm, thus in the next section, we review the existing consensus algorithms. 

\subsection{Consensus Algorithms}\label{sub-sec-consensus}
In this section, we discuss some of the well-known consensus algorithms in the literature and analyze their limitations in IoT.

Bitcoin is the first distributed cryptocurrency introduced in 2008 which employs Proof Of Work (POW) as the underlying consensus algorithm \cite{nakamoto2008bitcoin}. POW involves a computationally demanding, hard-to-solve, and easy-to-verify cryptographic puzzle which requires the miners, i.e., the vlaidators, to find a nonce value in a way that the hash of the block content along with nonce starts with a particular number of zeros. This, however, demands significant computational resources from the participating nodes. In recent years, particular mining devices known as ASIC miners are manufactured which  offer high hash rate. The difficulty, i.e., the number of leading zeros, of POW is dynamically adjusted to ensure that only one block an be mined during each 10 minutes. As POW difficulty increases, mining pools emerged where a group of miners work on a single block and share the revenue to enhance their chance of mining a block. Mining pools may    lead  to centralization as pools with large number of  participants  potentially may have  large portion of  mining power. \par
In POW only the miner that solves the puzzle and thus stores the next block is rewarded. This potentially wastes the resources of other miners that simultaneously worked on the same block.  The authors in  \cite{szalachowski2019strongchain} proposed a modified version of the POW that rewards the miners that partially solved the POW puzzle to provide further incentive for the nodes to participant in mining process.\par

Ethereum  \cite{wood2014ethereum}  proposes Proof of Stake (POS)   consensus algorithm where the mining power of the validators  is identified based on the amount of assets the validator  locked in the blockchain. The validators with more locked assets have more mining power which potentially increases the chance of such node to store the next blocks. POS significantly reduces the computational resource consumption of blockchain compared to POW, however, in IoT with large dominant companies, e.g., Google, POS might potentially lead to centralization.  \par

Proof Of Authority (POA) \cite{de2018pbft} is a consensus algorithm  which conceptually shares similarities with POS. In POA, the mining power of each validator comes from their identity rather than the amount of locked assets in POS. The validators are limited to a selected pre-defined nodes that are known to all participating nodes in the blockchain. This potentially raises privacy concern as the participants can track the revenue gained by each validator.  \par

Intel proposed a time-based consensus algorithm known as Proof of Elapsed Time (POET) \cite{chen2017security}.  POET is a leader-based consensus algorithm where the participants   choose a leader to store the next block.   The candidate validators generate a random waiting time and the validator with the shortest time is selected as the leader.    The random waiting time is generated in Trusted Execution Environment (TEE) in Intel CPUs which prevents against malicious validators that may claim to always have a short waiting time. Thus, POET requires all the participants to be equipped with Intel CPU  which is challenging in IoT with millions of heterogenous devices.  \par

In Federated Byzantine Algorithm (FBA) \cite{yoo2019formal}  each potential validator randomly selects a group of other validators and forms a quorum. The validators then share their quorums. If there exist intersections between quorums, the validator that is chosen by more validators is selected as the leader. In case no intersection can be found the network may take apart, forking occurs in the blockchain. FBA incurs high packet overhead  and processing delay as the validators has to broadcast the quorum information and analyze the quorum of other validators for each new block stored in the blockchain. \par 

The existing blockchain  consensus algorithms (some of which discussed above)  suffer from a number of challenges for IoT ecosystem which are discussed below: \par 

\begin{itemize}
	\item Throughput management: In a blockchain context, throughput is defined as the total number of transactions that can be stored in the blockchain per second. The Bitcoin throughput, for instance, is seven transactions per second.  IoT consists of millions of devices, SPs, and users that communicate  through transactions which potentially leads to millions of transactions which is far beyond the current throughput of the blockchains. Although new consensus algorithms improved the blockchain throughput, the ever increasing number of devices and services in IoT demands a self-scaling blockchain.  Thus, an IoT-friendly consensus algorithm is required to adjust the network throughput as the number of transactions increases. 
	
	\item Computational efficiency: In most of the existing blockchain instantiations, the validators attempt to append pending transactions in the blockchain simultaneously. The validtor that first follows the consensus algorithm rules wins the competition and thus can store block in the blockchain, while the resources spent by other validators is wasted as they shall start mining the next block.

	\item Delay:	In most of the existing consensus algorithms, mining transactions in the blockchain involves delay that is for the validators to run the consensus algorithm and reach agreement over the state of the blockchain. This delay increases as each node has to wait for a particular number of blocks to be chained in the current block before accepting a transaction. In IoT ecosystem, the transactions are employed to offer personalized services to the end-users, which demands  near real-time transaction processing time. As an example,  a smart home user cannot wait 30 seconds for the smart lock to open the door of the home. 
	
	\item Transaction retrieval: The participating nodes in the blockchain, in particular the validators, may need to retrieve a previously stored transaction, e.g., a validator may need to retrieve the previous transaction of the newly received transaction for verification. For this the validator has to search the blockchain database which in turn incurs delay and processing overhead. IoT users demand frequent access to their previously stored transactions which in turn amplifies the corresponding delay. 
	
	\item Resource consumption: The existing consensus algorithms consume significant computational, bandwidth, or storage resources of the validators. The resource consuming consensus algorithms aim to protect against double spending attack, where a malicious node spends the same coin twice. However,  IoT applications may not involve asset transmission and thus double spending may not be as relevant as it is for cryptocurrencies. 
\end{itemize}

In this paper, we propose tree-chain that provides a comprehensive solution to the aforementioned challenges and is discussed in details in Section \ref{Sec:HashCons}.

\section{Tree-Chain}\label{Sec:HashCons}

\begin{table}
	\caption{Definition of the abbreviations and indexes used in this paper.} 
	\centering
	\begin{tabular}{|c|c|}	\hline
		Notion & Meaning	\\ \hline
		PN\textsubscript{i} & Participating nodes in the blockchain  \\ \hline
		\textit{val\textsubscript{j}} & Blockchain validators  \\\hline
		\textit{t\textsubscript{i}} & Transaction generated by \textit{PN\textsubscript{i}} \\\hline
		\textit{pk\textsubscript{val}\textsuperscript{+}} & public key of node "validator" \\\hline
		$\ell \textsubscript{j}$ & A  blockchain ledger generated by \textit{val\textsubscript{j} }\\\hline
		$ \Delta$ & Epoch time to store blocks \\\hline
		$	\eth$ &  pre-setup time for consensus code formation\\\hline
		$\delta$ & time interval for which a transaction is valid \\\hline
		J & set of validators  \\\hline
	\end{tabular}
	
	\label{tab:devices}
\end{table}

\begin{figure}
	\begin{center}
		\includegraphics[width=10cm ,height=9cm ,keepaspectratio]{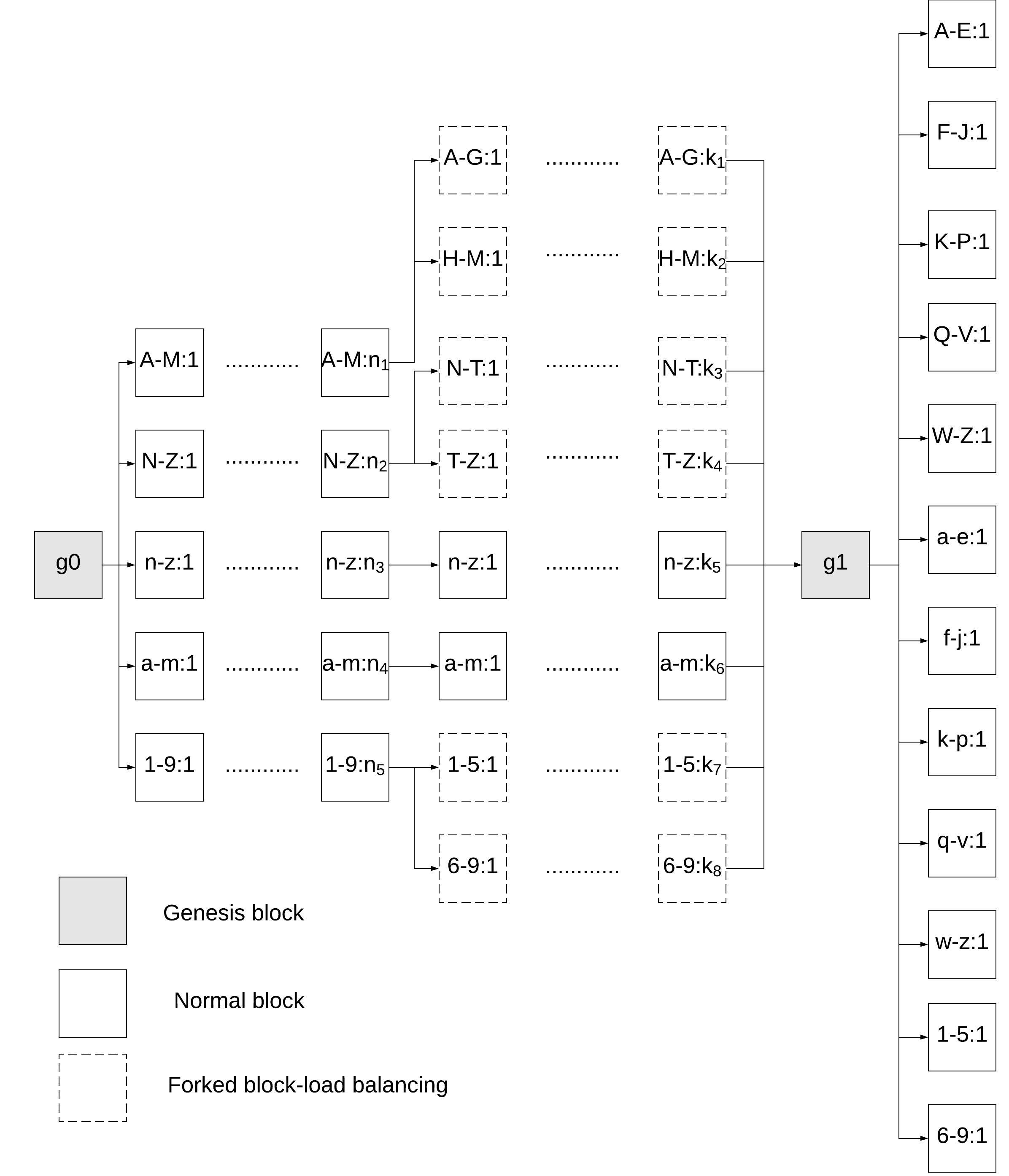}   
		\caption{An overview of Tree-chain.}
		\label{fig:HashCons}
	\end{center}
\end{figure}

This section outlines the details of Tree-Chain that addresses the limitations of the existing frameworks as outlined in Section \ref{sec:literature-review}. Table \ref{tab:devices} represents the list of abbreviations and indexes used in this paper. Each index, say index \textit{i}, refers to the varying index,  \textit{i\textsuperscript{f}} refers to the last character in the set, and \textit{I} refers to the set of indexes. Tree-chain introduces a load balancing algorithm (as discussed in Section \ref{subsubsection:loadbalancing}) that ensures self-scaling feature of the blockchain and thus addresses the throughput challenge. While all validators store the entire chain, the intuition of tree-chain is that the selection of validators for committing transactions and blocks to the ledger can be randomized at minimal computational cost, using the hash function outputs.  The validators commit transactions based on the most significant bits of the hash of the transaction which is referred to as consensus code. Each validator is randomly allocated to a particular consensus code  for an epoch period.  Thus, each transaction is committed to the ledger  only by one randomly selected validator which in turn increases efficiency. The consensus algorithm demands no extra computational or processing which in turn reduces the delay in storing new blocks to near real-time. Tree-chain is a non-linear  blockchain  structure (see Figure \ref{fig:chains}.b) where the transactions in each ledger share the same consensus code  which in turn  speeds up the  transaction retrial process.   As shown in Figure \ref{fig:HashCons},  each validator only commits  blocks in  a particular ledger for  a particular time-frame known as \textit{consensus period}, represented  as $ \Delta$.  The setup process for each $ \Delta$  takes an extra  	$	\eth$ that is the time taken to setup the consensus code and   is discussed later in this section. \par

At the beginning of $ \Delta$  the validators are allocated to a particular ledger based on their \textit{pk\textsuperscript{+}}. The allocation information is stored in a block known as the genesis block as shown with gray boxes in Figure \ref{fig:HashCons}. The validators chain their ledgers to the genesis block. Tree-chain can be considered as a leader selection algorithm where a leader is selected for a period of time to append transactions for a given consensus code. The usage of leadership algorithms in blockchain is not new. POET \cite{chen2017security} , POA \cite{de2018pbft}, and FBA  \cite{yoo2019formal} employ leadership algorithms where the leader eventually appends one single block in the blockchain. Tree-chain significantly reduces the packet and processing overhead for selecting a leader by extending the duration that a leader is valid, while ensuring  the randomness of the transactions that the leader can store.  At the end of each $ \Delta$ the validators are changed (see Section \ref{sub-sec-vali-reformation})  that enhances the randomization level and thus  protects against malicious validators  that may attempt to store fake transactions in the blockchain.   The number of  ledgers in the blockchain equals with the number of validators.  To achieve randomness among the potential validators and to protect the security of the ledger tree-chain introduces two randomization levels which are:\par

i)  Transaction level: The main aim of this level is to identify the validator in charge of storing  transactions  with a given  \textit{consensus code} value.  Assume t\textsubscript{i}  represents  a transaction generated by node \textit{i}.   $ h(t\textsubscript{i}) = \{ \beta \textsubscript{1}\beta \textsubscript{2}\beta \textsubscript{3}...\beta \textsubscript{k\textsuperscript{f}}  \alpha \textsubscript{k\textsuperscript{f}+1} ...\alpha \textsubscript{n\textsuperscript{f}}\} $ where \textit{h(t\textsubscript{i})} represents the hash of \textit{t\textsubscript{i}} and  \textit{n} is the size of the hash function output, \textit{k} is the size of the consensus code, and $ \alpha\textsubscript{n} \;\&\; \beta\textsubscript{k}  \in  \mathbb{H} $ where $ \mathbb{H}=\{0,1,2,3,...,9,a,b,c,...,z,A,B,C,...,Z\} $.\par

ii) Blockchain level: The main aim of this level is to identify the validator corresponding to a ledger based on  \textit{pk\textsubscript{val}\textsuperscript{+}}. The validator of each ledger collects transactions that start with a \textit{consensus code}, thus each ledger corresponds to a particular consensus code. Let's assume $ \forall j,  \; h(pk\textsubscript{j}\textsuperscript{+}) $ represents the hash of the \textit{pk\textsubscript{val}\textsuperscript{+}} of \textit{val\textsubscript{j}}.   $ h(pk\textsubscript{j}\textsuperscript{+}) = \{ \alpha \textsubscript{1}\alpha \textsubscript{2} \alpha \textsubscript{3}...\alpha \textsubscript{n\textsuperscript{f}}\}$.    The  validators run a randomization algorithm (see Section \ref{sub-sec-validator-selection}) that allocates a particular consensus code to each \textit{val\textsubscript{j}}. Each validator is responsible for  transaction hashes whose most significant characters matches its consensus code. In other words, a validator with consensus code $ \beta \textsubscript{1}\beta \textsubscript{2}\beta \textsubscript{3}...\beta \textsubscript{k}  $ is then  only responsible for collecting transactions (represented by \textit{t}) where h(t)= $ \beta \textsubscript{1}\beta \textsubscript{2}\beta \textsubscript{3}...\beta \textsubscript{k\textsuperscript{f}}\alpha \textsubscript{n\textsuperscript{f}-k\textsuperscript{f}} ... \alpha \textsubscript{n\textsuperscript{f}}$. \par 

Tree-chain consists of four main phases which are:\par 
\textbf{\textit{1)}} Validator selection\par  
\textbf{\textit{2)}} Block generation\par 
\textbf{\textit{3)}} Validator reformation \par 
\textit{\textbf{3) }}Load balancing\par 

A high-level overview of the steps involved in tree-chain is  outlined in Algorithm \ref{alg:tree-chain} and  details are discussed  in  the rest of this section:

\begin{algorithm}[tb!]
	\caption{Tree-chain}\label{alg:tree-chain}
	\begin{algorithmic}[1]
		\State  Send \textit{t\textsuperscript{vi}}\Comment{\textit{Validator Selection}}
		\State Calculate KWM 
		\State Form consensus code
		\State Inform validators of selected consensus code
		\If{KWM of \textit{j} is Max among validators }
		\State \textit{j} to generate genesis block
		\EndIf
		\State Collect  trans within consensus code  \Comment{Block generation}
		\If{pending-pool.size $>$ block.size \textbf{or} time $>$ block.time}
		\State Create the hash of trans in block
		\State Append new block to the ledger
		\State Broadcast block in the network
		\EndIf
		\If{time.now()=$\Delta\textsubscript{n}-\eth$}\Comment{Validator Reformation}
		\State Follow steps 1-4
		\EndIf
		\If{validator is overloaded}\Comment{Load Balancing}
		\State Broadcast new validator request to network
		\State Receive  \textit{t\textsuperscript{vi}}
		\State  Calculate KWM
		\State Select node with highest KWM as validator 
		\State Divide consensus code
		\EndIf
	\end{algorithmic}
\end{algorithm}

\subsection{Validator Selection}\label{sub-sec-validator-selection}
\begin{figure}
	\begin{center}
		\includegraphics[width=5cm ,height=5cm ,keepaspectratio]{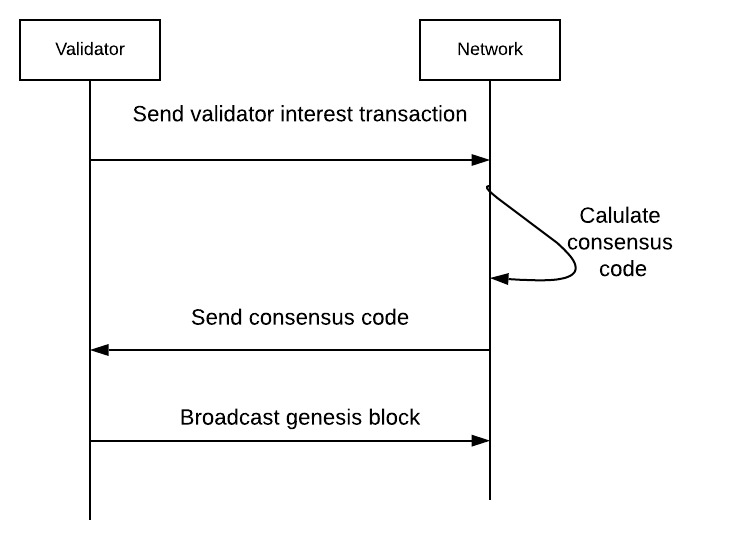}   
		\caption{The process of selecting validators}
		\label{fig:flowchart}
	\end{center}
\end{figure}

Tree-chain is a non-linear blockchain where a particular  validator appends blocks in its corresponding ledger for   each	$\Delta$. Each validator is allocated to a     consensus code  based on  their \textit{pk\textsubscript{val}\textsuperscript{+}}.  Recall that the setup of the consensus code take $\eth$ which is an additional time to $\Delta$. Thus, $\Delta\textsubscript{m}= \Delta\textsubscript{m} - \eth$, where $\Delta\textsubscript{m}$  is  round \textit{m} of   epoch time. Consensus code setup involves four main tasks, which are discussed in the next paragraph, each taking $\eth/4$ that is referred to as $ \eth\textsubscript{m}$ where \textit{m=1,2,3,4}. It is assumed that \textit{PN\textsubscript{i}} are time synchronized \cite{elson2002fine,mills1991internet}. The value of  $\Delta$ should be chosen by considering the end-to-end delay in the network to ensure that  transactions can be distributed during each $\delta$. Figure \ref{fig:flowchart} depicts a high level view of the process involved in selecting validators that is outlined in details in the rest of this section.\par

During $ \eth\textsubscript{1}$ \textit{PN\textsubscript{i}}s that are interested to function as validator, express their interest by generating a validator interest  transaction (represented as \textit{t\textsuperscript{vi}}) that is structured as $<t\_id,  pk, sign> $ where $t\_id$ is the transaction identifier that is the hash of the transaction content. \textit{pk} corresponds to the \textit{pk\textsubscript{i}\textsuperscript{+}} of the node that must be verifiable  through a Certificate Authority (CA). This  protects against sybil attack where a single node pretends to be multiple nodes by generating multiple \textit{t\textsuperscript{vi}}s (see Section \ref{sec:security-analysis}). \textit{sign} is the signature corresponding to \textit{pk\textsubscript{i}\textsuperscript{+}}. \textit{t\textsuperscript{vi}} is  broadcast to the network. Any \textit{t\textsuperscript{vi}} that is generated after  $ \eth\textsubscript{1}$  is discarded by the network.\par 

\begin{figure}
	\begin{center}
		\includegraphics[width=8cm ,height=9cm ,keepaspectratio]{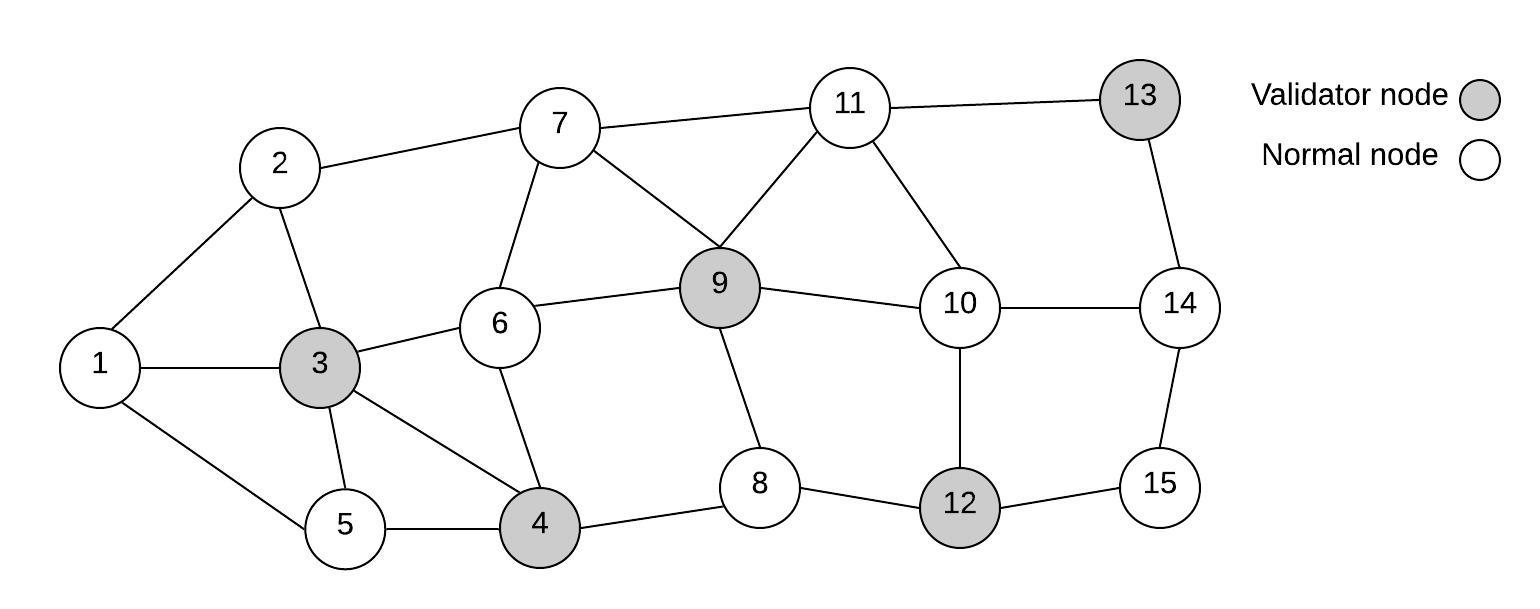}   
		\caption{Tree-chain network.}
		\label{fig:overlay}
	\end{center}
\end{figure}

During $ \eth\textsubscript{2}$ \textit{PN\textsubscript{i}}  receive \textit{t\textsuperscript{vi}} and verify it by verifying the \textit{pk} using CA and matching the \textit{sign} with \textit{pk}.   For each received  \textit{t\textsuperscript{vi}}, \textit{PN\textsubscript{i}} calculates a KWM as  $ KWM = \sum_{r=1}^{n} \omega(\alpha \textsubscript{r}) $ where $ \omega(\alpha \textsubscript{r})  $ is  a numerical weight corresponding to  each possible value of $ \alpha \textsubscript{n}$. Recall that  $ \alpha \textsubscript{n} \in  \mathbb{H}$. The weight corresponding to each $ \alpha \textsubscript{n}$ is extracted from a  \textit{Key Weight Metric (KWM)} dictionary, an example of which is shown in Table \ref{tab:KWMExample}. All \textit{PN\textsubscript{i}} apply the same KWM dictionary to ensure they all have the same view of KWM. The blockchain designers populate the KWM dictionary.  As an illustrative example, Table \ref{tab:tableexample} presents h(pk\textsubscript{j}\textsuperscript{+}), the corresponding KWM, and the allocated consensus code (discussed later in this section) for nodes shown in Figure \ref{fig:overlay}. To calculate the KWM, we employ Table \ref{tab:KWMExample} as KWM dictionary. \par

Recall that the consensus code is $ \beta \textsubscript{1}\beta \textsubscript{2}\beta \textsubscript{3}...\beta \textsubscript{k\textsuperscript{f}}  $  where  $  \beta \textsubscript{k} \in  \mathbb{H}$ and \textit{k} is the size of the consensus code. \textit{k} depends on the total number of received and verified \textit{t\textsuperscript{vi}}   during $ \eth\textsubscript{1}$. The value of \textit{k} should be chosen in a way that each \textit{PN\textsubscript{j}} is allocated  a unique  consensus code. As $ \beta \textsubscript{k} \in \mathbb{H} $, each $ \beta\textsubscript{k}$ can accommodate maximum of 62 validators. In case of 62 validators, each validator is allocated to one particular value in $\mathbb{H} $. If the number of validators exceeds 62, then \textit{$k>1$}. Note that we use this rule for explanation purposes, however, the allocation of the consensus code is a design choice.  \par 

\textit{PN\textsubscript{i}} create a descending ordered list of  the received \textit{t\textsuperscript{vi}} based on  KWM  value, represented as \textit{KWM\textsubscript{1}, KWM\textsubscript{2}, ... , KWM\textsubscript{j\textsuperscript{f}}} where \textit{KWM\textsubscript{1}} is the first in the list. \textit{PN\textsubscript{j}} corresponding to \textit{KWM\textsubscript{1}} is selected as the  validator  of the first range of the consensus code, represented by \textit{code\textsubscript{1}}. As an  example, consider the network shown in Figure \ref{fig:overlay}. The network consists of    5 validators thus  \textit{k=1}.   Table \ref{tab:tableexample} outlines the ID, PK, the KWM, and the consensus code range for each validator. KWM is measured based on the dictionary represented  in Table \ref{tab:KWMExample}.  As \textit{KWM\textsubscript{12}} is the highest value, 12 is allocated to the first consensus range (assuming that the consensus code range priority is numbers, uppercase letters and lowercase letters).  \textit{KWM\textsubscript{2}} is  selected as the backup validator for \textit{code\textsubscript{1}} that i) monitors the behavior of the main validator to detect any malicious activity, and ii) functions as backup in case the main validator is disconnected or leaves the network.  The consensus range is then allocated to the other validators based on KWM in a descending order. \par 

\begin{table}[h]
	\caption{Example of KWM dictionary.}\centering
	\begin{tabular}  { | p {2.5 cm} |   p {2 cm}|}
		\hline
		\textbf{ $ \alpha $}     & \textbf{$ \omega(\alpha)  $} \\\hline
		0-9 & 0-9  \\\hline
		A-Z & 11-36  \\\hline
		a-z & 	37-62 \\\hline
	\end{tabular}
	\label{tab:KWMExample}
\end{table}

\begin{table*}[h]
	\centering
	\caption{An example of consensus table based on Figure 1.}
	\begin{tabular}  { | p {2.5 cm} |   p {3 cm}| p {3 cm}| p {3 cm}|}
		\hline
		\textbf{ID }     & \textbf{PK}  & \textbf{KWM}  & \textbf{Consensus code range} \\\hline
		12& 	axqPe96aiwZjQ&482  &1-9 \\\hline
		3& aQfx12ijAtcTM &419  & A-M \\\hline
		4 & J94Vswa72liac&  356 & N-Z \\\hline
		13& 	Mq83V2mq62kEl& 341  & a-m\\\hline
		9& 	Rnah72Mec123a&  314 & n-z\\\hline

	\end{tabular}
	\label{tab:tableexample}
\end{table*}

During $ \eth\textsubscript{3}$  \textit{val\textsubscript{j}} sends its own consensus code range along with the total number of validators, i.e., \textit{j\textsuperscript{f}}, from its perspective to all validators. Each validator decides on the split in consensus code range by dividing the consensus code range  by  the total number of validators. This ensures that \textit{val\textsubscript{j}}s are consistent about the total number of validators and their corresponding consensus code range. \textit{val\textsubscript{j}}s that receive this packet reply with confirmation after checking the values. Any inconsistency is resolved  by considering the 66\% majority of \textit{val\textsubscript{j}}. This number is inspired from the  Byzantine algorithm \cite{yoo2019formal} and may vary depending on the application. \par 

During $ \eth\textsubscript{4}$ \textit{val\textsubscript{j}} with the highest KWM generates a block, also known as genesis block (\textit{b\textsubscript{gen}}).  Tree-chain differentiates between the first genesis block and the subsequent genesis blocks. The subsequent genesis blocks, e.g., \textit{g1} in Figure \ref{fig:HashCons}, are the last blocks in the previous $ \Delta$ that store the hash of the ledgers generated during $ \Delta$. Thus once the process of generating subsequent genesis blocks is started, the validators shall no longer generate a block. The genesis blocks are structured as follows: \par 

$ <Total\textsubscript{val}, <pk\textsuperscript{+}\textsubscript{vali\textsubscript{w}}, consensus\_ code\textsubscript{vali\textsubscript{w}}, sign\textsubscript{vali\textsubscript{w}}, \\hash\textsubscript{l\textsubscript{w} }>> $\par 
where  $  w = J_\Delta  \cup J_\Delta\textsubscript{-1} $ and $ J_\Delta$ represents the set of validators in round $\Delta$. \textit{Total\textsubscript{val}} is the total set of validators for the next $\Delta $. The next field is  a tuple  that includes \textit{pk\textsuperscript{+}}, consensus code range, signature and hash of the ledger for each \textit{val\textsubscript{w}}. hash\textsubscript{l\textsubscript{w}} is the hash of the ledger of blocks generated by \textit{val\textsubscript{w}} during the last $\Delta $, i.e., $\Delta-1$. For new validators, hash\textsubscript{l\textsubscript{w} }  is set to null, i.e., if $ val\textsubscript{w} \in J_\Delta \;\& \; \notin J_\Delta\textsubscript{-1} $ then $hash\textsubscript{l\textsubscript{w} } = null $. If a validator  in the previous round no longer wishes to function as validator, then its corresponding  consensus code is set as null, i.e.,  if $ val\textsubscript{w} \notin J_\Delta \;\& \; \in J_\Delta\textsubscript{-1}  $ then $consensus\_ code\textsubscript{vali\textsubscript{w}} = null $. Once \textit{val\textsubscript{w}} populates the genesis block with hash\textsubscript{l\textsubscript{w} }, it should no longer generate new blocks. \textit{b\textsubscript{gen}}  must  be signed by more than 66\% of the participants to be considered as a valid genesis block.   \par 

Following the outlined steps, \textit{PN\textsubscript{i}} agree on the validators of the next blocks. \textit{val\textsubscript{j}} starts storing new blocks once the new epoch, i.e.,  $\Delta$, starts. 

\subsection{Block generation} \label{sub-sec-block-generation}
Each \textit{val\textsubscript{j}} collects and verifies  transactions in its consensus range, as an example, \textit{val\textsubscript{3}} in Table \ref{tab:tableexample} stores all transactions where $ t.hash = {\beta,\alpha\textsubscript{1}, ...,\alpha\textsubscript{k\textsuperscript{f}-1}} $, where $ \beta \in {A,B,C, ...,M}$. Tree-chain enables two modes for generating new blocks: i) block size: where \textit{val\textsubscript{j}} generate a new block when size of the pending transactions, i.e., the transactions that are not yet stored in the blockchain, reaches a pre-defined value known as \textit{block.size}, and  ii) block time: where \textit{val\textsubscript{j}}  generate a new block at the end of particular time intervals, e.g., 10 seconds, known as \textit{block.interval}. In networks with low transaction load, one may consider generating blocks based on block time that will set an upper bound for the delay experienced by the users to store their transactions in the blockchain, while  in networks with significant number of transactions block size can be considered that standardizes the size of blocks in the ledger. Note that the same method applies to all validators.\par 

In tree-chain appending a new block to the ledger does not require  \textit{val\textsubscript{j}} to solve any puzzle or provide proof of X. Recall that tree-chain achieves randomization in two layers, which are blockchain and transaction layers, and thus eliminates the need for solving a puzzle before storing new blocks.  Thus, the upper-bound throughput for \textit{val\textsubscript{j}} is the speed at which \textit{val\textsubscript{j}} can collect transactions, verify them, and form new blocks which is relatively fast. Recall that tree-chain has a non-linear structure where each \textit{val\textsubscript{j}} chain blocks to its own ledger.  \par

To ensure that all \textit{val\textsubscript{j}}  have the same view of the overall state of the blockchain, $ <hash.ledger\textsubscript{1},hash.ledger\textsubscript{2},.., hash.ledger\textsubscript{j\textsuperscript{f}},merkle-tree> $ is appended to the block headers. Due to high speed in block generation, it is possible that  \textit{val\textsubscript{A}}   do not receive  a new block from other \textit{val\textsubscript{j}} between two blocks generated by \textit{val\textsubscript{A}}. In this case, \textit{val\textsubscript{A}}   sets the corresponding   $hash.ledger\textsubscript{j}$  as null to reduce the associated overheads. Storing the hash of all the ledgers in the blockchain increases the blockchain database size over time. Each \textit{val\textsubscript{j}} generates a merkle tree using $ <hash.ledger\textsubscript{1},hash.ledger\textsubscript{2},.., hash.ledger\textsubscript{j\textsuperscript{f}}>$ and stores the root of  the merkle tree in \textit{merkle-tree} field. At the end of $\Delta$, \textit{val\textsubscript{j}} removes $ <hash.ledger\textsubscript{1},hash.ledger\textsubscript{2},.., hash.ledger\textsubscript{j\textsuperscript{f}}>$ from the block headers while maintaining  $merkle-tree $. Tree-chain employs the method as proposed in \cite{dorri2019mof} that enables validators to remove the outlined fields while protecting the consistency of the chain. Note that the hash of each ledger is stored in \textit{b\textsubscript{gen}} that can later be used to validate transactions in a ledger. \par 

\textit{val\textsubscript{j}}  populates   \textit{pk\textsubscript{val\textsubscript{j}}\textsuperscript{+}} and the corresponding signature  on the block. \textit{pk\textsubscript{val\textsubscript{j}}\textsuperscript{+}} must be the same pk used during consensus code formation that ensures only the nodes that have been selected during consensus code formation generate blocks. The validator then broadcasts the block to the network. \par

Upon receipt of the new block, \textit{b\textsubscript{arv}}, \textit{val\textsubscript{j}} verifies the block by following the below steps: \par 

\begin{enumerate}
	\item Verify if the consensus code of the transactions in \textit{b\textsubscript{arv}} matches with the consensus code corresponding to the \textit{b\textsubscript{arv}}.generator using the \textit{pk\textsubscript{val\textsubscript{j}}\textsuperscript{+}} field in the block header.
	
	\item Verify the signature of \textit{b\textsubscript{arv}}  using the corresponding \textit{pk\textsubscript{val\textsubscript{j}}\textsuperscript{+}}.
	
	\item Verify the merkle tree of the ledger hashes in  \textit{b\textsubscript{arv}}  header.
	
	\item Verify if the transaction timestamp is  within  a particular time interval of the current time represented by $\delta$ that is defined by the blockchain designer. This is similar to expiry time in conventional blockchains.  This protects against malicious nodes that may attempt to double spend a transaction as discussed in Section \ref{sec:security-analysis}.
	
	\item If a transaction, say \textit{t\textsubscript{in}}, in \textit{b\textsubscript{arv}}  is spending the output of a previous transaction, say \textit{t\textsubscript{out}}, then  \textit{val\textsubscript{j}}   requests the previous transaction's validator  \textit{val\textsubscript{out}} to sign the transaction. Recall that in tree-chain \textit{val\textsubscript{j}}  store blocks simultaneously, thus a malicious node may attempt to conduct double spending attack, i.e., spend the same coin twice. Malicious node may generate two transactions spending the same coin to two users. The hash of the transactions and thus the validators are different  which  may lead to  both transactions to be stored in the blockchain. Tree-chain protects against this attack where \textit{val\textsubscript{j}} request the validator of \textit{t\textsubscript{out}}  to sign the same. Upon receipt of the request, the validator of \textit{t\textsubscript{out}} verifies the transaction, signs the request, and return it back to \textit{val\textsubscript{j}}. To ensure that the transaction is not double spent, the validator of \textit{t\textsubscript{out}}  sets a \textit{spent} flag for the transaction and add its signature. These fields are not included in the  calculation of  the block hash and thus do not affect the blockchain consistency. Similar to conventional blockchains, \textit{val\textsubscript{j}} maintains a list of unspent transactions to speed up the verification of transactions that spend unspent output.     We will further study the double spending attack in Section \ref{sec:security-analysis}. 
\end{enumerate}

The verification of the transaction may involve more processes depending on the blockchain application. \textit{val\textsubscript{j}} generate blocks as outlined above during $\Delta$. At the end of $\Delta$  the validators are reformed as outlined in the next section.

\subsection{Validator reformation }\label{sub-sec-vali-reformation}
In each $\Delta$ time, all validators reform the validators list which: i) enables new validators to join the validators' list, and ii) introduces a further layer of randomization as the consensus code corresponding to a validator changes in each $\Delta$. \par 

The reformation process starts at $ time.\Delta\textsubscript{n} - \eth $ where $time.\Delta\textsubscript{n}$  is the scheduled start time of the next epoch. Recall that $\eth$ allows  \textit{val\textsubscript{j}} to choose new validators in the background before $\Delta\textsubscript{n-1} $  is finished. This ensures that the end users do not experience delay for their transactions to be stored in the blockchain due to consensus code formation.  \par 

The process of consensus code reformation is the same as outlined in Section \ref{sub-sec-validator-selection}. In case \textit{val\textsubscript{j}} decides to continue its role for the next   $\Delta $  round, it must use  a new  \textit{pk\textsubscript{j}\textsuperscript{+}}. Using the same \textit{pk\textsubscript{j}\textsuperscript{+}} in $\Delta\textsubscript{n}  $ and $\Delta\textsubscript{n-1}  $  may lead to  the same consensus code allocation, which in turn impacts the validator randomization. To address this challenge,   $ \forall j ; \;   pk\textsuperscript{+}\textsubscript{j} \in  \Delta\textsubscript{n}    \neq pk\textsuperscript{+}\textsubscript{j} \in \Delta\textsubscript{n-1}   $. All \textit{t\textsuperscript{vi}} inconsistent with the outlined rule, are discarded.    The dynamic validator selection also provides a load balancing benefit for large-scale IoT networks, by incorporating additional validators when the load increases, as discussed below.\par

\subsection{Load Balancing}\label{subsubsection:loadbalancing}

As the number of transactions generated by \textit{PN\textsubscript{i}} increases, the number of transactions in each consensus code range also increases. Recall that the hash function output is random, thus there is no guarantee that the load is equally divided between validators. \textit{val\textsubscript{j}}, referred to as overloaded validator, \textit{val\textsubscript{over}}, in the rest of this section,  may be overwhelmed with a large number of transactions in its corresponding consensus code. In such case, \textit{val\textsubscript{over}} can request  new validator to join by  sending  a  validator request to  \textit{PN\textsubscript{i}}. The interested \textit{PN\textsubscript{i}} reply  by sending a \textit{t\textsuperscript{vi}}(see Section \ref{sub-sec-validator-selection}). \textit{val\textsubscript{over}}  follows the same steps as in Section \ref{sub-sec-validator-selection} and selects the node with the highest KWM as new validator.  \textit{val\textsubscript{over}}  divides the corresponding consensus code range in two and allocates each range to one validator.  As the consensus code range is divided, the ledger is also is divided by creating a new fork that is used by the new validator to store blocks.  As an example,  in Figure \ref{fig:HashCons} "A-M" range is divided into two leading to two new ledgers "A-G" and "H-M". The first range is allocated to the overloaded validator, and the second is allocated to the new validator. The load balancing algorithm assists \textit{val\textsubscript{over}} to reduce the processing overhead till the next $\Delta$. Thus, there is a tradeoff for \textit{val\textsubscript{over}} to consider the delay in adding a new validator and the time  till the end of the current   $\Delta$. We leave the detailed discussion for our future work. \par

As the output of the hash function is random, it cannot be guaranteed that the load is divided equally between the validators. In case after load balancing a validator is still overloaded, the process is repeated for that particular validator. Utilizing the outlined process, tree-chain achieves self-scaling feature.

\section{Security Analysis}\label{sec:security-analysis}
In this section, we analyze   the security of tree-chain against various attacks. We assume that adversary, represented as  \textit{adv\textsubscript{k}} where $ k \subset I  \; \& \; k>=1$, can sniff communications,  discard transactions (\textit{t\textsubscript{i}}) and blocks (\textit{b\textsubscript{j}}), and create false transactions (\textit{t\textsubscript{fa}}) and blocks (\textit{b\textsubscript{fa}}). We assume that standard encryption methods are in use and cannot be compromised by the adversary. We study four different attacks below: 
\par 
\textbf{\textit{Double Spending Attack: }} Assume \textit{adv\textsubscript{k}} owns a digital asset $ \kappa\textsubscript{k}  $ according to a transaction \textit{t\textsubscript{in}}. The transfer of $ \kappa\textsubscript{k}  $ from  \textit{adv\textsubscript{k}} to a \textit{PN\textsubscript{i}} is represented as $\kappa\textsubscript{k} \Rightarrow \kappa\textsubscript{i}$.  In this attack, that is known as double spending attack,    $ \kappa\textsubscript{k} \Rightarrow \kappa\textsubscript{x}  \; \& \; \kappa\textsubscript{k} \Rightarrow \kappa\textsubscript{y}  \; $where $\; x,y \in I$. \par 
\textit{Defense:}  Tree-chain introduces a layered defense that makes it impossible to conduct double spending. Recall that tree-chain is designed for IoT applications where the asset transmission is not as common as cryptocurrencies thus double spending may not apply for all transactions. However, we study this attack to provide a comprehensive study on tree-chain security.   \textit{adv\textsubscript{k}}  may conduct this attack using one of the following methods: \par 
i) \textit{adv\textsubscript{k}}    generates and broadcasts transactions corresponding to $ \kappa\textsubscript{k} \Rightarrow \kappa\textsubscript{x} $ and $\kappa\textsubscript{k} \Rightarrow \kappa\textsubscript{y}$ simultaneously. Given that the content of these transactions, and thus the corresponding hash, varies, different \textit{val\textsubscript{j}}  may attempt to store transactions simultaneously which potentially leads to a successful double spending. Similar to the conventional blockchains,   when \textit{val\textsubscript{j}} receives a transaction that transfers an asset, say \textit{t\textsubscript{in}}, it checks the blockchain to verify if the output of   \textit{t\textsubscript{in}} has been spent. As \textit{adv\textsubscript{k}} generate the transactions simultaneously, the above verification will pass and both  \textit{val\textsubscript{j}}  store transactions.  \par 

Tree-chain protects against this in two layers. As outlined in Section \ref{sub-sec-block-generation}, before verifying \textit{t\textsubscript{in}}, \textit{val\textsubscript{j}} requests \textit{val\textsubscript{m}} to sign the transaction where $ m \in J  $  and the hash of the transaction to be spent, represented as \textit{t\textsubscript{out}}, falls in the consensus code range of \textit{m}. Upon receipt of the request  \textit{val\textsubscript{m}} marks  \textit{t\textsubscript{out}} as spent by setting a flag. Later if \textit{t\textsubscript{out}}  is used as input in another transaction,   \textit{val\textsubscript{m}} will receive another request from \textit{val\textsubscript{j}}. Given the flag is set as spent, \textit{val\textsubscript{m}}  informs \textit{val\textsubscript{j}} that \textit{t\textsubscript{out}} has already been spent. \par 

Recall that tree-chain achieves two levels of randomization in blockchain and transaction levels. However,  it still might be possible for   \textit{adv\textsubscript{k}} to control \textit{val\textsubscript{m}}. If so, \textit{adv\textsubscript{k}} confirms both transactions. The second protection layer relies on the distributed nature of the blockchain. As all transactions are broadcast, \textit{val\textsubscript{j}} will eventually receive the blocks containing the double spent transactions and thus can detect the double spending during block verification. In such case, the malicious behavior of \textit{adv\textsubscript{k}}, i.e., \textit{val\textsubscript{m}},  is reported to the network. The network utilizes the double spent transactions as evidence and agree on the malicious behavior of \textit{adv\textsubscript{k}}. Thus, \textit{adv\textsubscript{k}} is removed from the validators list. To prevent \textit{adv\textsubscript{k}} to re-join the network, the CA prevents issuing new \textit{pk\textsuperscript{+}} to the \textit{adv\textsubscript{k}}.  \par

ii) \textit{adv\textsubscript{k}} attempts to generate two transactions corresponding to $ \kappa\textsubscript{k} \Rightarrow \kappa\textsubscript{x}  \; \& \; \kappa\textsubscript{k} \Rightarrow \kappa\textsubscript{y}  \; $where$ \; x,y \in PN$ in a way that hash of the transactions falls within the consensus code range of \textit{adv\textsubscript{k}}. Hash function output is completely random and thus \textit{adv\textsubscript{k}} cannot manipulate it. \textit{adv\textsubscript{k}} can only conduct brute forcing by changing the transaction values in a way that the final hash falls in a particular range.  A transaction is structured as $<T\_ID,timestamp, input, output, pk, sign>$. \textit{adv\textsubscript{k}} may only change \textit{timestamp} and \textit{pk} to conduct brute force. Recall from Section \ref{Sec:HashCons} that the transactions in a block must be generated within a particular time range of the block generation time which limits the possible range of values for \textit{timestamp}. Creating new \textit{pk} and thus a new \textit{sign} incurs significant computational overhead on \textit{adv\textsubscript{k}}.\par 

Depending on the number of \textit{val\textsubscript{j}}, there is always a chance for a successful double spending as discussed above in method (ii). We studied the  time taken to conduct the attack (details of the implementation are outlined in Section \ref{sec:performance}).   \textit{adv\textsubscript{k}} continues changing the timestamp  of the transaction  until the hash of the transaction falls within the consensus code associated  with  \textit{adv\textsubscript{k}}. Table \ref{tab:double-spending-attack} represents the  implementation results which are the average of 10 runs of the algorithm. We assumed there is one adversary in the network.\par

\begin{table}
	
	\centering
	\caption{Processing time to conduct double spending attack.}\label{tab:double-spending-attack}
	\begin{tabular}{|p{1cm}|p{2cm}|}
		\hline
		\textbf{\textit{j}}  & \textbf{\textit{Time(s) }}\\\hline
		10 &  19 \\\hline
		20 &  52.9  \\\hline
		50 & 265.1  \\\hline
		100  & 1714.6 \\\hline
		
	\end{tabular}
	
\end{table}

It is expected that a large number of validators will participate in Tree-chain as storing new blocks does not involve solving any puzzle or spending resources. With the large number of validators, the consensus code range increases which  reduces the probability of a successful double spending attack as proven by the results discussed above.  As outlined earlier in this section, tree-chain provides two protection layers, and thus the double spending attack will eventually be detected by the participating nodes in the network. \par

\textbf{\textit{Denial of Service Attack:}}  \textit{adv\textsubscript{k}} selectively store transactions in its corresponding consensus code range which potentially impacts the services received by  \textit{PN\textsubscript{i}} whose transactions fall in the consensus code corresponding to \textit{adv\textsubscript{k}}.\par 

\textit{Defense:} The impact of this attack depends on the size of \textit{J}, as with more validators the impact of this attack is limited as fewer transactions will fall in a particular consensus code range.  Tree-chain protects against this attack benefiting from the distributed nature of the blockchain. \textit{val\textsubscript{j}} monitor the  cumulative number of transactions generated within a particular consensus code range and the number of such transactions actually stored by the the corresponding  validator.  In case that the difference between two values reaches a particular threshold, defined by the blockchain designer, \textit{val\textsubscript{j}} choose a new validator for the corresponding consensus code by following the same process as outlined in Section \ref{sub-sec-validator-selection}. \par

\textbf{\textit{Sybil Attack:}}  \textit{adv\textsubscript{k}} attempts to control a broader  consensus code range by pretending to be   multiple  \textit{PN\textsubscript{i}} by advertising multiple  \textit{pk\textsuperscript{+}}s. By increasing the controlled consensus code range,  \textit{adv\textsubscript{k}}  aims to  increase the probability of a successful double spending attack.

\textit{Defense:} Tree-chain requires a \textit{pk\textsuperscript{+}} used during consensus code formation step to be certified by a CA. The latter may require the requesters to either provide documents to identify them-selves, or pay a specific amount. Thus, employing multiple  \textit{pk\textsuperscript{+}} potentially increases \textit{adv\textsubscript{k}} cost.  Recall from Section \ref{Sec:HashCons} that tree-chain introduces two levels of randomization which are in blockchain and transaction level. Even if \textit{adv\textsubscript{k}} succeeds in controlling a larger percentage of \textit{val\textsubscript{j}}, the transaction level randomization cannot be controlled.  \par

\textbf{\textit{Node Isolation Attack:}} \textit{adv\textsubscript{k}} attempts  to isolate a group of \textit{val\textsubscript{j}}  by dropping packets to or from them.  This may  decrease the number of \textit{val\textsubscript{j}} and thus lead to a larger  consensus code range  for \textit{adv\textsubscript{k}}  that enables them to conduct double spending attack.

\textit{Defense:}  Similar to other existing blockchains, tree-chain provides a number of entry points which are  the nodes whose address is publicly announced so the new nodes can join the network. A malicious entry point may connect new nodes to a set of  malicious nodes that isolate new nodes from tree-chain. New \textit{PN\textsubscript{i}}    may use multiple entry points which in turn connects them to a broader range of nodes in tree-chain and thus mitigates the impact of this attack. As studied earlier in this section,  tree-chain provides multiple layers of security and thus increasing the consensus code range will not guarantee a successful attack.

\textit{\textbf{Summary of security analyses:}} In this section, we analyzed  the security of  tree-chain against a range of malicious behaviors. It is proven that the security of  tree-chain largely relies on two levels of randomization in blockchain and transaction level which make it complicated and resource consuming for an attacker to conduct attack. Tree-chain is designed for large scale networks such as IoT and thus the number of \textit{val} is expected to be large which in turn limits the impact of malicious behaviors. 
\par

Having discussed the security, we next  discuss the fault tolerance of tree-chain. \par 

\textbf{\textit{Fault tolerance:}}   Fault tolerance reflects  the resilience of an architecture against failure of \textit{PN\textsubscript{i}}. As \textit{val\textsubscript{j}} store new blocks,  their failure may impact the fault tolerance of tree-chain. Recall from  Section \ref{sub-sec-block-generation} that  there exists a backup validator for each consensus code that covers the failed validator for the consensus  code.  If both \textit{val\textsubscript{j}} and the backup validator fail, the transactions in the corresponding consensus range  will no longer be stored in the blockchain which is detected by other \textit{val\textsubscript{j}} as outlined earlier in this section. The validator that stored the genesis block in this $ \Delta $ will initiate the process to select a new validator as outlined in Section \ref{sub-sec-validator-selection}.
Thus, failure of a \textit{val\textsubscript{j}} has limited impact on the transactions that fall within that particular consensus code range. To improve the fault tolerance of tree-chain multiple \textit{val\textsubscript{j}} may collaboratively generate transactions with a particular consensus code range which we leave for future work.\par

\section{Performance Evaluation}\label{sec:performance}
In this section we study the performance of tree-chain.  As discussed in   Section \ref{Sec:HashCons}, tree-chain incorporates fundamental  changes  to  conventional blockchains and thus we were unable to use the existing simulation environments, such as Hyperledger Fabric \cite{HyperledgerFabric}, to study the performance of tree-chain. We implemented full functions of  tree-chain    using  Java programming language. To prove that tree-chain is runnable by low resource available IoT devices, we studied the performance of tree-chain on Raspberry Pi 2. The presented results are the average of 10 runs of the algorithm. We studied the following metrics: \par 

\begin{itemize}
	\item Consensus code formation processing time: This is the time taken for each \textit{val\textsubscript{j}}  to follow  $\eth\textsubscript{1} - \eth\textsubscript{4} $     as outlined in Section \ref{sub-sec-validator-selection}. We disregarded  the communication delay as it depends on the network setting and is not impacted by tree-chain design.
	
	\item New block generation processing time: This is the time taken for each \textit{val\textsubscript{j}}  to follow steps in \ref{sub-sec-block-generation} to generate a block and append it to the blockchain. 
	
	\item Load balancing: This metric evaluates the impact of the load balancing algorithm by studying the processing overhead  on  \textit{val\textsubscript{j}}   before and after running the load balancing algorithm. 
	
	\item Double spending: This metric  evaluates the processing time incurred to protect against double spending attack. Recall that in tree-chain \textit{val\textsubscript{j}} must connect to the validator of the transaction used as the input of the current transaction which increases the delay in verifying transactions.
	
	\item Transaction retrieval:  This metric evaluates the processing time incurred to retrieve a transaction from the blockchain. In conventional blockchains all blocks are chained in a single ledger, while in tree-chain transactions are chained in different ledgers based on the consensus code. Thus, to retrieve a particular transaction, the nodes shall just search the relevant ledger. 
\end{itemize}

\begin{figure}
	\begin{center}
		\includegraphics[width=7cm ,height=9cm ,keepaspectratio]{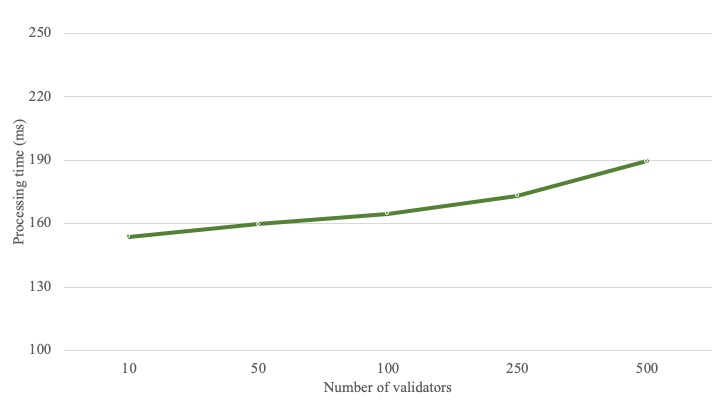}   
		\caption{Evaluation of the processing overhead during consensus code formation.}
		\label{fig:consensusCodeFormation}
	\end{center}
\end{figure}

\textit{Consensus code formation processing time:} To evaluate this metric, we increase the number of potential validators, i.e., the nodes that generate \textit{t\textsuperscript{vi}},  from 10 to 500.  Evident from the results shown in  Figure \ref{fig:consensusCodeFormation}  the processing time increases from around 153 ms to 190 ms which is negligible delay. Tree-chain    only demands the \textit{val\textsubscript{j}}  to calculate a value using the hash output, which is already in the received \textit{t\textsuperscript{vi}} and thus does not incur significant processing overhead. \par 

The formation of the consensus code also incurs packet overhead on the validators. As outlined in Figure \ref{fig:flowchart}, each validator  broadcasts two packets during the consensus code formation. Assume that  the size  of a packet is $ \psi$ and each \textit{val}  receives a packet only once.  The cumulative packet overhead for these packets will be (2$\psi$j). The validator with the highest KWM broadcasts the genesis block at the end of consensus formation. Assume the size of the genesis block is $\Psi$, thus the total packet overhead in consensus formation in each round is: \textit{Packet overhead = (2$\psi$j)+$\Psi$}.\par 

As outlined earlier in Section \ref{Sec:HashCons}, the  value of $\Delta$ is defined by the blockchain designers. The latter shall consider the trade-off between  the overheads associated with small   $\Delta$ and the security risks with large $\Delta$. Smaller $\Delta$ requires the validators to run validator  reconfiguration algorithm more frequently (see Section \ref{sub-sec-vali-reformation}) which incurs packet and processing overhead as studied above, while longer $\Delta$ increases the chance of a  double spending attack as studied in Section \ref{sec:security-analysis}. Recall that tree-chain ensures that double spending can eventually be detected.\par

\begin{figure}
	\begin{center}
		\includegraphics[width=7cm ,height=9cm ,keepaspectratio]{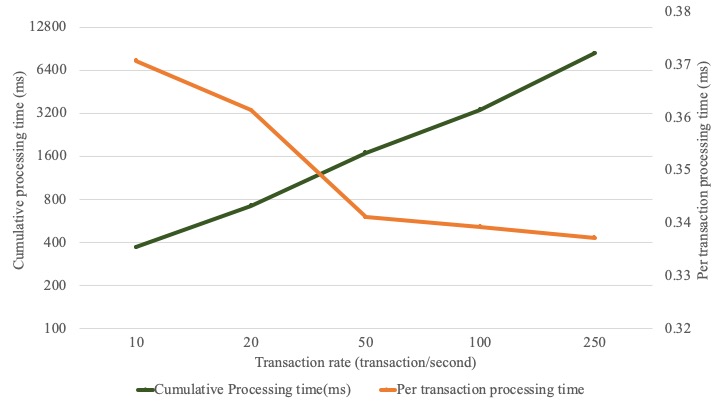}   
		\caption{Evaluation of the processing overhead for forming new blocks.}
		\label{fig:processingtime}
	\end{center}
\end{figure}

\textit{New block generation processing time:} Recall that tree-chain does not demand  \textit{val\textsubscript{j}} to solve any puzzle before storing a new block, thus storing a new block simply involves collecting transactions in the consensus code range associated with the validator, forming a new block once the size or time reached, and appending it to the blockchain (see Section \ref{sub-sec-block-generation}).  Figure \ref{fig:processingtime} outlines the implementation results on the processing time to store new blocks.  The horizontal axis refers to the transaction rate in which the \textit{PN\textsubscript{i}} generate transactions. The left axis  refers to the cumulative processing time for generating new blocks while the right axis refers to the  average processing overhead to  store a single transaction. In our implementation setting \textit{i=100}.  We assume there are 10 validators in the network. Block size is 10 transactions per block.

As evident from the results, the processing time required to append a new block is around 370 ms when \textit{PN\textsubscript{i}} generate 10 transactions per second which is close to real-time. Note that this processing time is the cumulative processing time on a single validator to store all blocks to accommodate transactions which in case of 10 transactions/second is 1000 transactions.  Note that each validator commits only transactions with specific consensus code. As the transaction rate increases from 10 to 250, the processing time increases from around 370 ms to 8400 ms. This shows that tree-chain can store transactions in  near real-time. As evident from the results, by increasing the transaction rate, the processing time for each transaction reduces in tree-chain. This is because the higher transaction rate allows transaction pools to reach the block size more quickly, which further highlights the scalability of tree-chain. \par

As evident from the results shown in Figure \ref{fig:processingtime}, tree-chain achieves a fast block generation time. IoT nodes generate millions of transactions and blocks. This potentially increases the bandwidth consumption of the blockchain. As tree-chain block generation is fast, the number of blocks broadcast in the network increases that is inherent from the IoT. Thus, \textit{val\textsubscript{j}} may experience  bandwidth limitations that  potentially limits the number of transactions that can reach to \textit{val\textsubscript{j}} and  may impact the delay in storing transactions and   thus the upper bound throughput of tree-chain. The bandwidth limitation is beyond the scope of this paper, however, technologies such as 5G or 6G can be used to increase the bandwidth of \textit{val\textsubscript{j}}.

\begin{figure}
	\begin{center}
		\includegraphics[width=7cm ,height=8cm ,keepaspectratio]{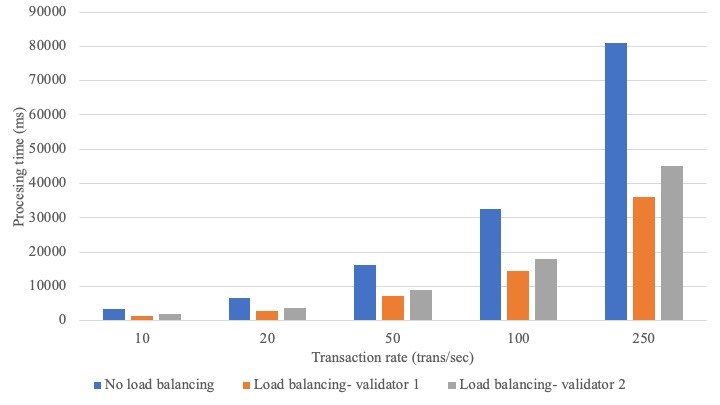}   
		\caption{Evaluating the impact of  load balancing on the processing overhead on \textit{val\textsubscript{j}}.}
		\label{fig:loadbalancing}
	\end{center}
\end{figure}

\textit{Load balancing:}  Recall from Section \ref{subsubsection:loadbalancing} that tree-chain enables overloaded  \textit{val\textsubscript{j}} to add new validators and thus reduce the processing overhead.  Figure  \ref{fig:loadbalancing} shows the implementation results for evaluation of the processing overhead on  \textit{val\textsubscript{j}} before and after applying load balancing algorithm.  Based on the results shown in Figure \ref{fig:loadbalancing}, by adding a new validator, the processing overhead  almost halves. Recall that after load-balancing each validator is allocated to a new consensus code range. Due to the randomness of the hash function, the processing overhead is similar but not equal  between the two validators. \par

\begin{figure}
	\begin{center}
		\includegraphics[width=6cm ,height=6cm ,keepaspectratio]{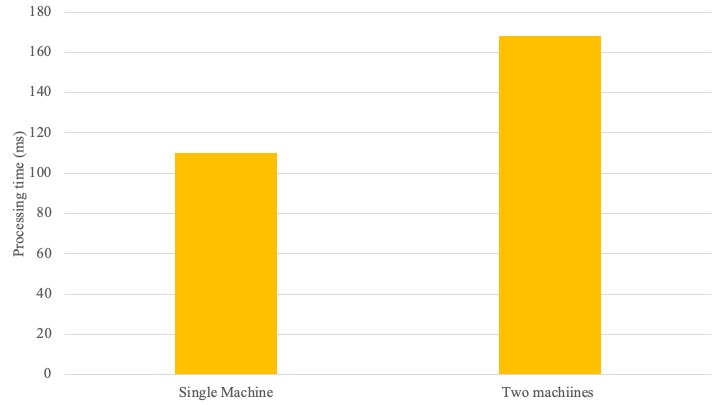}   
		\caption{Evaluation of the incurred  processing overhead for double spending verification.  }
		\label{fig:doublespending-veri}
	\end{center}
\end{figure}

\textit{Double spending:} Recall from Section \ref{Sec:HashCons} that  to protect against double spending, when \textit{val\textsubscript{j}}   receives a transaction  that spends the output of a previous transaction, it has to request \textit{val}   corresponding to the consensus code range of the previous transaction to sign the transaction and verify that the transaction is not double spent. In this part of evaluation, we study the processing time incurred to verify the transactions.There are two factors that impact the delay involved in verifying a transaction which are processing delay incurred on the verifier, i.e., \textit{val} of the previous transaction, and the communication delay. To measure the processing delay incurred on the verifier, we studied the delay when both \textit{vals} are in the same machine, i.e., a Raspberry Pi,  which eliminates the communication delay.  To show the impact of the communication delay, we measured the delay when each \textit{val}  runs on a separate Raspberry Pi. The Pi devices  are in the same network and connected through a router.  The implementation results are shown in Figure \ref{fig:doublespending-veri}.  The delay in verifying a transaction is increased by about 50 ms which is the communication delay. However, the processing overhead incurred on the verifier is not changing. The verification of the transaction involves verifying the signature which is a resource consuming task on Pi devices, thus this delay reduces in higher resource available devices. \par 

\begin{figure}
	\begin{center}
		\includegraphics[width=7cm ,height=9cm ,keepaspectratio]{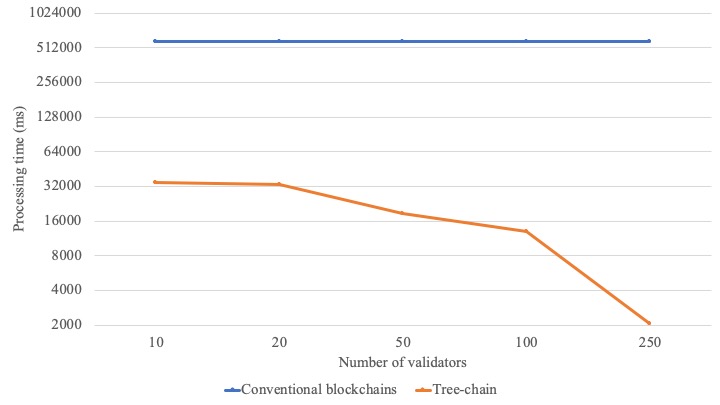}   
		\caption{Evaluation of the processing time to retrieve a transaction.  }
		\label{fig:transaction-riterieval}
	\end{center}
\end{figure}

\textit{Transaction retrieval:} In an IoT blockchain, it is highly common for \textit{PN\textsubscript{i}} to retrieve  a previously stored transaction, e.g., in a supply chain scenario the participants need to audit different steps of the product by retrieving transactions in the blockchain. In tree-chain transactions with particular consensus code are chained in a separate ledger which in turn speeds up the transaction retrieval as the query can run over the shorter ledger branch in tree-chain. Figure \ref{fig:transaction-riterieval} outlines the processing overhead incurred on a node to retrieve a transaction.   100 million transactions are stored in the blockchain and the blockchain database size is 110 GB. As Raspberry Pis have limited storage space, we employed  a Macbook Pro 15 to study the performance.  The processing overhead is around 587000 ms in conventional blockchains which is not impacted by the number of validators in the network, i.e., \textit{j}, as all \textit{val\textsubscript{j}} use a single chain to store transactions. In tree-chain, the processing overhead starts from 34500 ms when \textit{j=10}. This value decreases as new validators join the network and reaches 2000 ms when \textit{j=250}. This  is the result of fewer transactions being stored in a single ledger which in turn reduces the number of transactions to be searched. 

Based on the implementation results outlined in this section, tree-chain reduces the processing overhead on the validators to store new blocks and manage the blockchain.  

\section{Discussion and Future Research Directions}\label{sec:discussions}

\subsection{Permissioned vs Permissionless}

Depending on the permissions  and level of trust among \textit{PN\textsubscript{i}},  blockchains are categorized as permissioned and permissionless. In the former category,  a  trusted  node authorizes \textit{PN\textsubscript{i}}  to access the blockchain and  \textit{PN\textsubscript{i}}   have different read/write permissions, e.g., only selected nodes can  funciton as \textit{val\textsubscript{j}}. A permissionless blockchain, however, does not rely on any trusted node and \textit{PN\textsubscript{i}}  have equal control on the blockchain. In this paper we studied tree-chain in a permissionless setting as the absence of the trusted party leads to higher complexity and resource consumption in reaching consensus. However, tree-chain can be applied in permissioned blockchain with minor changes.

\subsection{Double Spending}
Tree-chain is designed for IoT ecosystems where transactions reflect communications between devices and some transactions may involve transferring an asset. As discussed in Section \ref{sec:security-analysis} to protect double spending, tree-chain requires the validators of transaction \textit{t\textsubscript{in}} that spends the output of transaction \textit{t\textsubscript{out}}  to ask the validator corresponding to  \textit{t\textsubscript{out}} to sign \textit{t\textsubscript{in}}. This provides an extra layer of security, however, adds overhead and increases the delay. Another interesting future research  direction is to optimize tree-chain for cryptocurrency. This requires \textit{val\textsubscript{j}} to maintain a single ledger, i.e., a linear ledger, to protect against double spending. Recall that tree-chain does not require \textit{val\textsubscript{j}} to solve any puzzle before storing a new block which speeds up the block generation rate, which in turn makes reaching agreement over the chain of ledgers in the blockchain challenging as the number of forks increases (see Section \ref{Sec:HashCons}).

\subsection{Replication}
In tree-chain,  each   \textit{val\textsubscript{j}}  is dedicated to a consensus code. Failure of  \textit{val\textsubscript{j}}   may lead to service disruption for transactions that fall within the consensus code range of the failed \textit{val\textsubscript{j}}. As discussed in Section \ref{sec:security-analysis} each \textit{val\textsubscript{j}} monitors \textit{val} responsible for the next consensus code. To further improve the  fault tolerance, multiple  \textit{val\textsubscript{j}} can work on the same consensus code which conceptually is similar to having multiple replications. Distributing transactions and synchronization among the replicas remain major challenges. 

\subsection{State Pinning }

Tree-chain defines genesis blocks which are stored at the end of each $\Delta$. The genesis block can be employed to pin the state of the blockchain. In \cite{robinson2019anonymous} the authors discussed state pinning in Ethereum. Pinning the state of the blockchain can be employed to  reduce the blockchain storage overhead, e.g., some \textit{val\textsubscript{j}} may decide not to maintain the full history of the blockchain and thus can only store blocks after the pinned state.

\section{Conclusion}\label{sec:conclusion}

This paper proposed tree-chain, a scalable fast blockchain optimized for IoT applications. Tree-chain incorporates a consensus algorithm that does not demand the validators to solve any puzzle or provide proof of x before storing a new block. The randomization among the validators is achieved by relying on the hash function outputs. Two randomization levels are introduced which are i) transaction level where the validator of each transaction is defined randomly based on the most significant bits of the hash of the transaction (known as consensus code), and ii) blockchain level where each validator is dedicated to store transactions with particular consensus code. Tree-chain introduces a load-balancing algorithm that enables  the overloaded validators  to involve new validators and thus ensures self-scaling feature of the blockchain. The implementation results prove tree-chain incurs low processing overhead and is runnable by low resource IoT devices.   Tree-chain will enable new fast blockchain applications in more resource-constrained scenarios such as IoT.

\bibliographystyle{IEEEtran}
\bibliography{IEEEabrv,bare_jrnl_compsoc}

\begin{thebibliography}{10}
\providecommand{\url}[1]{#1}
\csname url@samestyle\endcsname
\providecommand{\newblock}{\relax}
\providecommand{\bibinfo}[2]{#2}
\providecommand{\BIBentrySTDinterwordspacing}{\spaceskip=0pt\relax}
\providecommand{\BIBentryALTinterwordstretchfactor}{4}
\providecommand{\BIBentryALTinterwordspacing}{\spaceskip=\fontdimen2\font plus
\BIBentryALTinterwordstretchfactor\fontdimen3\font minus
  \fontdimen4\font\relax}
\providecommand{\BIBforeignlanguage}[2]{{%
\expandafter\ifx\csname l@#1\endcsname\relax
\typeout{** WARNING: IEEEtran.bst: No hyphenation pattern has been}%
\typeout{** loaded for the language `#1'. Using the pattern for}%
\typeout{** the default language instead.}%
\else
\language=\csname l@#1\endcsname
\fi
#2}}
\providecommand{\BIBdecl}{\relax}
\BIBdecl

\bibitem{atlam2018blockchain}
H.~F. Atlam, A.~Alenezi, M.~O. Alassafi, and G.~Wills, ``Blockchain with
  internet of things: Benefits, challenges, and future directions,''
  \emph{International Journal of Intelligent Systems and Applications},
  vol.~10, no.~6, pp. 40--48, 2018.

\bibitem{christidis2016blockchains}
K.~Christidis and M.~Devetsikiotis, ``Blockchains and smart contracts for the
  internet of things,'' \emph{Ieee Access}, vol.~4, pp. 2292--2303, 2016.

\bibitem{alphand2018iotchain}
O.~Alphand, M.~Amoretti, T.~Claeys, S.~Dall'Asta, A.~Duda, G.~Ferrari,
  F.~Rousseau, B.~Tourancheau, L.~Veltri, and F.~Zanichelli, ``Iotchain: A
  blockchain security architecture for the internet of things,'' in \emph{2018
  IEEE Wireless Communications and Networking Conference (WCNC)}.\hskip 1em
  plus 0.5em minus 0.4em\relax IEEE, 2018, pp. 1--6.

\bibitem{dorri2019lsb}
A.~Dorri, S.~S. Kanhere, R.~Jurdak, and P.~Gauravaram, ``Lsb: A lightweight
  scalable blockchain for iot security and anonymity,'' \emph{Journal of
  Parallel and Distributed Computing}, vol. 134, pp. 180--197, 2019.

\bibitem{ramachandran2018blockchain}
G.~S. Ramachandran and B.~Krishnamachari, ``Blockchain for the iot:
  Opportunities and challenges,'' \emph{arXiv preprint arXiv:1805.02818}, 2018.

\bibitem{qiu2018dynamic}
H.~Qiu, M.~Qiu, G.~Memmi, Z.~Ming, and M.~Liu, ``A dynamic scalable blockchain
  based communication architecture for iot,'' in \emph{International Conference
  on Smart Blockchain}.\hskip 1em plus 0.5em minus 0.4em\relax Springer, 2018,
  pp. 159--166.

\bibitem{ma2019privacy}
M.~Ma, G.~Shi, and F.~Li, ``Privacy-oriented blockchain-based distributed key
  management architecture for hierarchical access control in the iot
  scenario,'' \emph{IEEE Access}, vol.~7, pp. 34\,045--34\,059, 2019.

\bibitem{dorri2017blockchain}
A.~Dorri, M.~Steger, S.~S. Kanhere, and R.~Jurdak, ``Blockchain: A distributed
  solution to automotive security and privacy,'' \emph{IEEE Communications
  Magazine}, vol.~55, no.~12, pp. 119--125, 2017.

\bibitem{mengelkamp2018blockchain}
E.~Mengelkamp, B.~Notheisen, C.~Beer, D.~Dauer, and C.~Weinhardt, ``A
  blockchain-based smart grid: towards sustainable local energy markets,''
  \emph{Computer Science-Research and Development}, vol.~33, no. 1-2, pp.
  207--214, 2018.

\bibitem{lee2017blockchain}
B.~Lee and J.-H. Lee, ``Blockchain-based secure firmware update for embedded
  devices in an internet of things environment,'' \emph{The Journal of
  Supercomputing}, vol.~73, no.~3, pp. 1152--1167, 2017.

\bibitem{jo2018hybrid}
B.~W. Jo, R.~M.~A. Khan, and Y.-S. Lee, ``Hybrid blockchain and
  internet-of-things network for underground structure health monitoring,''
  \emph{Sensors}, vol.~18, no.~12, p. 4268, 2018.

\bibitem{wood2014ethereum}
G.~Wood \emph{et~al.}, ``Ethereum: A secure decentralised generalised
  transaction ledger,'' \emph{Ethereum project yellow paper}, vol. 151, no.
  2014, pp. 1--32, 2014.

\bibitem{Hyperledger}
\BIBentryALTinterwordspacing
``Hyperledger.'' [Online]. Available: \url{https://www.hyperledger.org}
\BIBentrySTDinterwordspacing

\bibitem{HyperledgerFabric}
\BIBentryALTinterwordspacing
``Hyperledger fabric.'' [Online]. Available:
  \url{https://www.hyperledger.org/projects/fabric}
\BIBentrySTDinterwordspacing

\bibitem{tomescu2017catena}
A.~Tomescu and S.~Devadas, ``Catena: Efficient non-equivocation via bitcoin,''
  in \emph{2017 IEEE Symposium on Security and Privacy (SP)}.\hskip 1em plus
  0.5em minus 0.4em\relax IEEE, 2017, pp. 393--409.

\bibitem{liu2019mathsf}
Y.~Liu, K.~Wang, Y.~Lin, and W.~Xu, ``Lightchain: A lightweight blockchain
  system for industrial internet of things,'' \emph{IEEE Transactions on
  Industrial Informatics}, vol.~15, no.~6, pp. 3571--3581, 2019.

\bibitem{luu2016secure}
L.~Luu, V.~Narayanan, C.~Zheng, K.~Baweja, S.~Gilbert, and P.~Saxena, ``A
  secure sharding protocol for open blockchains,'' in \emph{Proceedings of the
  2016 ACM SIGSAC Conference on Computer and Communications Security}, 2016,
  pp. 17--30.

\bibitem{nakamoto2008bitcoin}
S.~Nakamoto \emph{et~al.}, ``Bitcoin: A peer-to-peer electronic cash system,''
  2008.

\bibitem{szalachowski2019strongchain}
P.~Szalachowski, D.~Reijsbergen, I.~Homoliak, and S.~Sun, ``Strongchain:
  Transparent and collaborative proof-of-work consensus,'' \emph{arXiv preprint
  arXiv:1905.09655}, 2019.

\bibitem{de2018pbft}
S.~De~Angelis, L.~Aniello, R.~Baldoni, F.~Lombardi, A.~Margheri, and
  V.~Sassone, ``Pbft vs proof-of-authority: Applying the cap theorem to
  permissioned blockchain,'' 2018.

\bibitem{chen2017security}
L.~Chen, L.~Xu, N.~Shah, Z.~Gao, Y.~Lu, and W.~Shi, ``On security analysis of
  proof-of-elapsed-time (poet),'' in \emph{International Symposium on
  Stabilization, Safety, and Security of Distributed Systems}.\hskip 1em plus
  0.5em minus 0.4em\relax Springer, 2017, pp. 282--297.

\bibitem{yoo2019formal}
J.~Yoo, Y.~Jung, D.~Shin, M.~Bae, and E.~Jee, ``Formal modeling and
  verification of a federated byzantine agreement algorithm for blockchain
  platforms,'' in \emph{2019 IEEE International Workshop on Blockchain Oriented
  Software Engineering (IWBOSE)}.\hskip 1em plus 0.5em minus 0.4em\relax IEEE,
  2019, pp. 11--21.

\bibitem{elson2002fine}
J.~Elson, L.~Girod, and D.~Estrin, ``Fine-grained network time synchronization
  using reference broadcasts,'' \emph{ACM SIGOPS Operating Systems Review},
  vol.~36, no.~SI, pp. 147--163, 2002.

\bibitem{mills1991internet}
D.~L. Mills, ``Internet time synchronization: the network time protocol,''
  \emph{IEEE Transactions on communications}, vol.~39, no.~10, pp. 1482--1493,
  1991.

\bibitem{dorri2019mof}
A.~Dorri, S.~S. Kanhere, and R.~Jurdak, ``Mof-bc: A memory optimized and
  flexible blockchain for large scale networks,'' \emph{Future Generation
  Computer Systems}, vol.~92, pp. 357--373, 2019.

\bibitem{robinson2019anonymous}
P.~Robinson and J.~Brainard, ``Anonymous state pinning for private
  blockchains,'' in \emph{2019 18th IEEE International Conference On Trust,
  Security And Privacy In Computing And Communications/13th IEEE International
  Conference On Big Data Science And Engineering (TrustCom/BigDataSE)}.\hskip
  1em plus 0.5em minus 0.4em\relax IEEE, 2019, pp. 827--834.

\end{thebibliography}

\end{document}